\documentclass[pre,twocolumn,superscriptaddress,showpacs]{revtex4}
\usepackage{amsmath,amssymb,dsfont,graphicx,amssymb}
\usepackage[utf8x]{inputenc}
\usepackage[T1]{fontenc}
\usepackage[english]{babel}

\begin{document}

\title{Synchronization in networks with multiple interaction layers}
\author{Charo I. \surname{del Genio}}\email{C.I.del-Genio@warwick.ac.uk}
	\affiliation{School of Life Sciences, University of Warwick, Coventry, CV4 7AL, UK}
\author{Jesús \surname{Gómez-Gardeñes}}
	\affiliation{Departamento de Física de la Materia Condensada, University of Zaragoza, 50009 Zaragoza, Spain}
	\affiliation{Institute for Biocomputation and Physics of Complex Systems (BIFI), University of Zaragoza, 50018 Zaragoza, Spain}
\author{Ivan \surname{Bonamassa}}
	\affiliation{Department of Physics, Bar-Ilan University, 52900 Ramat Gan, Israel}
\author{Stefano Boccaletti}
	\affiliation{CNR--Istituto dei Sistemi Complessi, Via Madonna del Piano, 10, 50019 Sesto Fiorentino, Italy}
	\affiliation{Embassy of Italy in Israel, 25 Hamered Street, 68125 Tel Aviv, Israel}
\date{\today}

\begin{abstract}
The structure of many real-world systems
is best captured by networks consisting of several
interaction layers. Understanding how a multi-layered
structure of connections affects the synchronization properties
of dynamical systems evolving on top of it is a highly relevant endeavour
in mathematics and physics, and has potential
applications to several societally relevant topics,
such as power grids engineering and neural dynamics.
We propose a general framework to assess stability of the synchronized
state in networks with multiple interaction layers, deriving
a necessary condition that generalizes the Master Stability Function
approach. We validate our method applying it to a network
of Rössler oscillators with a double layer of interactions,  and show that highly
rich phenomenology emerges. This includes cases where
the stability of synchronization can be induced
even if both layers would have individually induced unstable synchrony, an effect genuinely
due to the true multi-layer structure of the
interactions amongst the units in the network.
\end{abstract}


\maketitle

\section{Introduction}
Network theory~\cite{Str001,AlB002,New003,DoM003,Ben004,Boc006,Cal007,New010,CoH010}
has proved a fertile ground for the modeling of a multitude of complex systems.
One of the main appeals of this approach lies in its power to identify universal
properties in the structure of connections amongst the elementary units of a system~\cite{WaS998,BaA999,DGM008}.
In turn, this enables researchers to make quantitative predictions about the collective
organization of a system at different length scales, ranging from the microscopic
to the global scale~\cite{Gui005,For010,del13,Pei14,Wil14,Tre15,New15}.

As networks often support dynamical processes, the interplay between structure
and the unfolding of collective phenomena has been the subject of numerous studies~\cite{BBV008,BaB013,GBB016}.
In fact, many relevant processes and their associated emergent phenomena, such
as social dynamics~\cite{CFL009}, epidemic spreading~\cite{Pas015}, synchronization~\cite{Boc002},
and controllability~\cite{LSB011}, have been proved to depend significantly on
the complexity of the underlying interaction backbone.
Synchronization of systems of dynamical units is a particularly
noteworthy topic, since synchronized states are at the core of
the development of many coordinated tasks in natural and engineered
systems~\cite{Pik001,Str003,Man004}. Thus, in the past two decades,
considerable attention has been paid to shed light on the role
that network structure plays on the onset and stability of synchronized states~\cite{PeC998,Lag000,Bar002,Nis003,Bel004,Hwa005,Cha005,Mot005,Zhou006,Lod007,JGG011,Bil14,del015}.

In the last years, however, the limitations of the simple network paradigm
have become increasingly evident, as the unprecedented availability of large
data sets with ever-higher resolution level has revealed that real-world
systems can be seldom described by an isolated network.
Several works have proved that mutual interactions between different complex
systems cause the emergence of networks composed by multiple layers~\cite{Boc014,Kiv014,Lee015,Bia015}.
This way, nodes can be coupled according to different kinds of ties so that
each of these interaction types defines an interaction layer.
Examples of multilayer systems include social networks, in which individual people are linked and
affiliated by different types of relations~\cite{Sze010}, mobility networks,
in which individual nodes may be served by different means of transport~\cite{Cardillo,Hal014},
and neural networks, in which the constituent neurons interact over chemical
and ionic channels~\cite{Adh011}.
Multi-layer networks have thus become the natural framework to investigate
new collective properties arising from the interconnection of different
systems~\cite{Rad013,JGG15}. The multi-layer studies of processes such as
percolation~\cite{Bul010,Son012,Gao012,BiD014,Bax016}, epidemics spreading~\cite{Men012,Gran013,Buono014,Sanz014},
controllability~\cite{MAB016}, evolutionary games~\cite{JGG012,Wang014,Mata015,Wang015}
and diffusion~\cite{Gom013} have all evidenced a very different phenomenology
from the one found on mono-layer structures. For example, while isolated
scale-free networks are robust against random failures of nodes or edges~\cite{Alb000},
interdependent ones are instead very fragile~\cite{Dan016}. Nonetheless,
the interplay between multi-layer structure and dynamics remains, under several
aspects, still unexplored and, in particular, the study of synchronization
is still in its infancy~\cite{Agu015,Zha015,Sev015,Gambuzza15}.

Here, we present a general theory that fills this gap, and generalizes the celebrated
Master Stability Function (MSF) approach in complex networks~\cite{PeC998} to the realm
of multi-layer complex systems. Our aim is to provide a full mathematical framework that
allows one to evaluate the stability of a globally synchronized state for non-linear
dynamical systems evolving in networks with multiple layers of interactions. To do this, we perform a linear stability analysis
of the fully synchronized state of the interacting systems,
and exploit the spectral properties of the graph Laplacians of each layer. The final result
is a system of coupled linear ordinary differential equations for the evolution of the
displacements of the network from its synchronized state.
Our setting does not require (nor assume) special conditions
concerning the structure of each single layer, except that
the network is undirected and that the local and interaction dynamics
are described by continuous and differentiable functions.
Because of this, the
evolutionary differential equations are non-variational.
We validate our predictions in a network of chaotic Rössler oscillators
with two layers of interactions featuring different topologies. We show that, even in this simple case, there is the possibility
of inducing the overall stability of the complete synchronization manifold in regions
of the phase diagram where each layer, taken individually, is known to be unstable.

\section{Results}
\subsection{The model}
\begin{figure}[b]
 \centering
\includegraphics[width=0.4\textwidth]{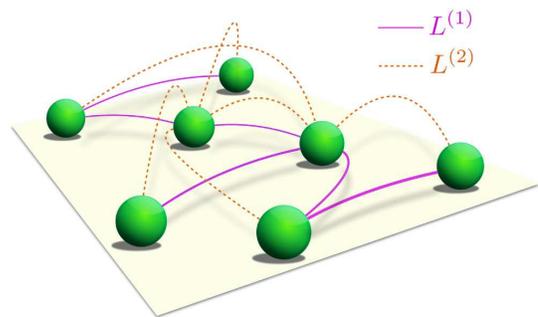}
\caption{\label{duplex}Schematic representation of
a network with two layers of interaction. The two layers
(corresponding here to solid violet and dashed orange links, respectively)
are made of links of different type for the same nodes, such as different
means of transport between two cities, or chemical
and electric connections between neurons. Note that
the layers are fully independent, in that they are described by two
different Laplacians $\mathbf L^{(1)}$ and $\mathbf L^{(2)}$, so that the presence
of a connection between two nodes in one layer does
not affect their connection status in the other.}
\end{figure}

From the structural point of view, we consider a network
composed of $N$ nodes which interact via $M$ different layers of connections, each layer
having in general different links and representing a different kind of
interactions among the units (see Fig.~\ref{duplex} for
a schematic illustration of the case of $M=2$ layers
and $N=7$ nodes).
Notice that in our setting the nodes interacting
in each layer are literally the same elements. Node~$i$ in layer~1
is precisely the same node as node~$i$ in layer~2, 3, or~$M$.
This contrasts with other works in which there is a one-to-one
correspondence between nodes in different layers, but these
represent potentially different states.
The weights of the connections between
nodes in layer $\alpha$ ($\alpha=1,\dotsc,M$) are given by
the elements of the matrix $\mathbf W^{\left(\alpha\right)}$,
which is, therefore, the adjacency matrix of a weighted graph.
The sum $q_i^\alpha=\sum_{j=1}^NW_{i,j}^{\left(\alpha\right)}$ ($i=1,\dotsc,N$) of
the weights of all the interactions of node $i$ in layer
$\alpha$ is the strength of the node in that layer.

Regarding the dynamics, each node represents a $d$-dimensional dynamical system.
Thus, the state of node $i$ is described by a vector $\mathbf{x}_i$
with $d$ components. The local dynamics of the nodes is captured
by a set of differential equations of the form
\begin{equation*}
\dot{\mathbf x}_i=\mathbf F\left(\mathbf{x}_i\right)\;,
\end{equation*}
where the dot indicates time derivative and $\mathbf F$
is an arbitrary $C^1$-vector field. Similarly,
the interaction in layer $\alpha$ is described
by a continuous and differentiable vector field $\mathbf H_{\alpha}$
(different, in general, from layer to layer),
possibly weighted by a layer-dependent coupling constant
$\sigma_\alpha$. We assume that the interactions
between node $i$ and node $j$ are diffusive, i.e., that
for each layer in which they are connected,
their coupling depends on the difference between $\mathbf H_{\alpha}$
evaluated on $\mathbf{x}_j$ and $\mathbf{x}_i$. Then, the
dynamics of the whole system is described by the following set of equations:
\begin{equation}\label{eomsys}
 \dot{\mathbf x}_i=\mathbf F\left(\mathbf{x}_i\right)-\sum_{\alpha=1}^M\sigma_\alpha\sum_{j=1}^NL_{i,j}^{\left(\alpha\right)}\mathbf H_{\alpha}\left(\mathbf{x}_j\right)\:,
\end{equation}
where $\mathbf L^{\left(\alpha\right)}$
is the graph Laplacian of layer $\alpha$, whose elements
are:
\begin{equation}\label{lapldef}
 L_{i,j}^{\left(\alpha\right)} = \begin{cases}
                                  q_i^\alpha &\quad\text{if }i=j\;,\\
                                  -W_{i,j}^{\left(\alpha\right)} &\quad\text{otherwise}\;.
                                 \end{cases}
\end{equation}

Let us note that our treatment of this setting is valid for
all possible choices of $\mathbf F$ and $\mathbf H_{\alpha}$, so long as
they are $C^1$, and for any particular undirected structure of the layers. This stands
in contrast to other approaches to the study of the same equation
set~(\ref{eomsys}) proposed in prior works (and termed as dynamical
hyper-networks), which, even though based on ingenious techniques such as simultaneous
block-diagonalization, can be applied only to special cases like commuting Laplacians, un-weighted
and fully connected layers, and non-diffusive coupling~\cite{Sor012},
or cannot guarantee to always provide a satisfactory solution~\cite{Irv012}.

\subsection{Stability of complete synchronization in networks with multiple layers of interactions}
We are interested in assessing the stability of synchronized states, which
means determining whether a system eventually returns to the synchronized solution after
a  perturbation. For further details of the following derivations we refer to Materials
and Methods.

First let us note that, since the Laplacians are
zero-row-sum matrices, they all have a null eigenvalue, with corresponding eigenvector
$N^{-1/2}\left(1,1,\dotsc,1\right)^{\mathrm T}$, where~T indicates transposition.
This means that the general system of equations~(\ref{eomsys}) always
admits an invariant solution $\mathbf{S}\equiv\{\mathbf{x}_i(t)=\mathbf{s}(t),\,\forall\,i=1,2,\dots,N\}$,
which defines the complete
synchronization manifold in $\mathbb{R}^{dN}$.

As one does not need a very strong forcing to destroy synchronization in an unstable state, we aim at
predicting the behavior of the system when the perturbation is small. Then, we first linearize Eqs.~(\ref{eomsys})
around the synchronized manifold $\mathbf{S}$ obtaining the equations ruling the evolution of the local and global
synchronization errors
$\delta\mathbf{x}_i\equiv\mathbf{x}_i-\mathbf s$ and
$\delta\mathbf X\equiv\left(\delta\mathbf{x}_1,\delta\mathbf{x}_2,\dotsc,\delta\mathbf{x}_N\right)^\mathrm{T}$:
\begin{equation}\label{linglob}
 \delta\dot{\mathbf X}=\left(\mathds 1\otimes J\mathbf F\left(\mathbf s\right)-\sum_{\alpha=1}^M\sigma_\alpha\mathbf L^{\left(\alpha\right)}\otimes J\mathbf H_{\alpha}\left(\mathbf s\right)\right)\delta\mathbf X\:,
\end{equation}
where $\mathds 1$ is the $N$-dimensional identity matrix,
$\otimes$ denotes the Kronecker product, and $J$ is the Jacobian
operator.

Second, we spectrally decompose $\delta\mathbf X$ in the equation above,
and project it onto the basis defined by the eigenvectors of
one of the layers. The particular choice of layer is completely arbitrary,
as the eigenvectors of the Laplacians of each layer
form $M$ equivalent bases of $\mathbb{R}^{N}$. In the following, to fix
the ideas, we operate this projection onto the eigenvectors of $\mathbf L^{\left(1\right)}$.
After some algebra, the system of equations~(\ref{linglob}) can be expressed as:
\begin{multline}\label{mainsystem}
 \dot{\boldsymbol\eta}_{j} = \left(J\mathbf F\left(\mathbf s\right)-\sigma_1\lambda_j^{(1)}J\mathbf{H}_1\left(\mathbf s\right)\right)\boldsymbol\eta_{j}+\\
 -\sum_{\alpha=2}^M\sigma_\alpha\sum_{k=2}^N\sum_{r=2}^N\lambda_r^{(\alpha)}\Gamma_{r,k}^{(\alpha)}\Gamma_{r,j}^{(\alpha)}J\mathbf H_{\alpha}\left(\mathbf s\right)\boldsymbol\eta_{k}\:,
\end{multline}
for $j=2,\dotsc,N$, where $\boldsymbol\eta_{j}$ is the vector
coefficient of the eigendecomposition of $\delta\mathbf X$,
$\lambda_r^{(\alpha)}$ is the $r$th eigenvalue
of the Laplacian of layer $\alpha$, sorted in
non-decreasing order, and we have
put ${\boldsymbol\Gamma}^{(\alpha)}\equiv\mathbf{V^{(\alpha)}}^\mathrm{T}\mathbf{V}^{(1)}$,
in which $\mathbf{V^{(\alpha)}}$ indicates
the matrix of eigenvectors of the Laplacian of
layer $\alpha$. Note that to obtain
this result, one must ensure that the Laplacian
eigenvectors of each layer are orthonormal, a
choice that is always possible because all the
Laplacians are real symmetric matrices. Thus,
the sums run
from~2 rather than~1 because the first eigenvalue
of the Laplacian, corresponding to $r=1$, is always~0
for all layers, and the first eigenvector, to which
all others are orthogonal, is common to all layers.
Equation~\ref{mainsystem} is notable
in that it includes prior results about systems
with commuting Laplacians as a special case. In
fact, if the Laplacians commute they can be simultaneously
diagonalized by a common basis of eigenvectors.
Thus, in this case, $\mathbf{V}^{(\alpha)}=\mathbf{V}^{(1)}\equiv\mathbf V$
for all $\alpha$. In turn, this implies that
$\boldsymbol\Gamma^{(\alpha)}=\mathds 1$ for
all $\alpha$, and Eq.~\ref{mainsystem} becomes
\begin{equation*}
 \begin{split}
   \dot{\boldsymbol\eta}_{j} &= \left(J\mathbf F\left(\mathbf s\right)-\sigma_1\lambda_j^{(1)}J\mathbf{H}_1\left(\mathbf s\right)\right)\boldsymbol\eta_{j}+\\
   & \quad -\sum_{\alpha=2}^M\sigma_\alpha\sum_{k=2}^N\sum_{r=2}^N\lambda_r^{(\alpha)}\delta_{r,k}\delta_{r,j}J\mathbf H_{\alpha}\left(\mathbf s\right)\boldsymbol\eta_{k}\\
   &= \left(J\mathbf F\left(\mathbf s\right)-\sum_{\alpha=1}^M\sigma_\alpha\lambda_j^{(\alpha)}J\mathbf H_{\alpha}\left(\mathbf s\right)\right)\boldsymbol\eta_{j}\:,\\
 \end{split}
\end{equation*}
recovering an $M$-parameter variational form as in \cite{Sor012}.

Notice that the stability of the synchronized state is
completely specified by the maximum conditional Lyapunov
exponent $\Lambda$, corresponding to the variation
of the norm of $\boldsymbol\Omega\equiv\left(\boldsymbol\eta_2,\dotsc,\boldsymbol\eta_N\right)$.
In fact, since $\boldsymbol\Omega$ will evolve on average as
$\left|\boldsymbol\Omega\right|\left(t\right)\sim \exp\left(\Lambda t\right)$, the fully synchronized state will be stable
against small perturbations only if $\Lambda<0$.

\subsection{Case study: networks of Rössler oscillators}
To illustrate the predictive power of the framework
described above, we apply it to a network of
identical Rössler oscillators, with two layers of connections. Note that our method is fully
general, and it can be applied to systems composed
by any number of layers and containing oscillators
of any dimensionality $d$. The particular choice of
$M=2$  and $d=3$ for our example allows us to
study a complex phenomenology, while retaining ease
of illustration. The dynamics of the Rössler oscillators
is described by $\dot{\mathbf x}=\left(-y-z,x+ay,b+\left(x-c\right)z\right)^\mathrm{T}$,
where we have put $x\equiv x_1$, $y\equiv x_2$ and $z\equiv x_3$.
The parameters are fixed to the values $a=0.2$, $b=0.2$ and $c=9$,
which ensure that the local dynamics of each node is chaotic.

Considering each layer of connections individually, it is known that the choice
of the function $\mathbf H$ allows (for an ensemble of networked Rössler oscillators)  the selection of one of
the three classes of stability (see Materials and
Methods for more details), which are:
\begin{itemize}
\item[I:] $\mathbf H\left(\mathbf x\right)=\left(0,0,z\right)$, for which synchronization is always unstable.
\item[II:] $\mathbf H\left(\mathbf x\right)=\left(0,y,0\right)$, for which synchronization is stable only for $\sigma_{\alpha}\lambda^{\alpha}_2<0.1445$.
\item[III:] $\mathbf H\left(\mathbf x\right)=\left(x,0,0\right)$ for which synchronization is stable only for ${}^{0.181}/{}_{\lambda^{\alpha}_2}<\sigma_{\alpha}<{}^{4.615}/{}_{\lambda^{\alpha}_N}$.
\end{itemize}

Because of the double layer structure, one can now combine together
different classes of stability in the two layers, studying how one
affects the other and identifying new stability conditions arising
from the different choices. In the following, we consider three combinations, namely:
\begin{itemize}
 \item \textbf{Case 1:} Layer~1 in class~I and layer~2 in class~II, i.e.,
 $\mathbf{H}_1\left(\mathbf x\right)=\left(0,0,z\right)$ and
 $\mathbf{H}_2\left(\mathbf x\right)=\left(0,y,0\right)$.
 \item \textbf{Case 2:} Layer~1 in class~I and layer~2 in class~III, i.e.,
 $\mathbf{H}_1\left(\mathbf x\right)=\left(0,0,z\right)$ and
 $\mathbf{H}_2\left(\mathbf x\right)=\left(x,0,0\right)$.
 \item \textbf{Case 3:} Layer~1 in class~II and layer~2 in class~III, i.e.,
 $\mathbf{H}_1\left(\mathbf x\right)=\left(0,y,0\right)$ and
 $\mathbf{H}_2\left(\mathbf x\right)=\left(x,0,0\right)$.
\end{itemize}

As for the choices of the Laplacians $\mathbf L^{\left(1,2\right)}$, we consider three possible combinations: (\emph{i})
both layers as Erd\H{o}s-Rényi networks of equal mean degree (ER-ER); (\emph{ii})
both layers as scale-free networks with power-law exponent~3 (SF-SF); and (\emph{iii})
layer~1 as Erd\H{o}s-Rényi and layer~2 as scale-free (ER-SF). In all cases,
the graphs are generated using the algorithm in Ref.~\cite{Gom006},
which allows a continuous interpolation between scale-free and
Erd\H{o}s-Rényi structures (see Materials and Methods for details).
Therefore, in the following we will consider~9 possible scenarios,
i.e., the three combinations of stability classes for each of the
three combinations of layer structures.
\begin{figure}[t]
\centering
\includegraphics[width=0.35\textwidth]{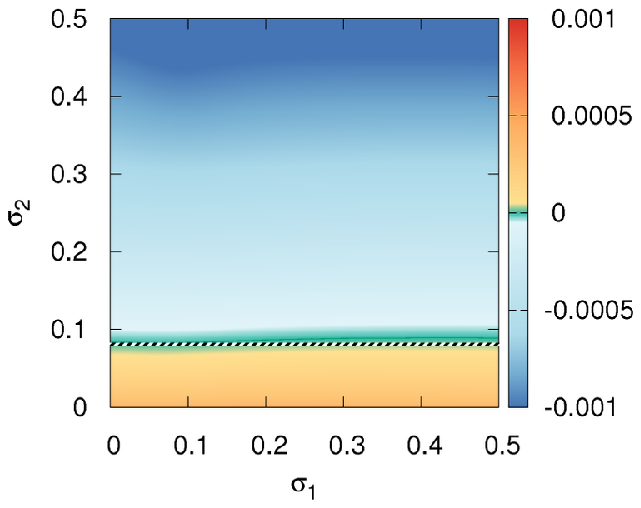}
\includegraphics[width=0.35\textwidth]{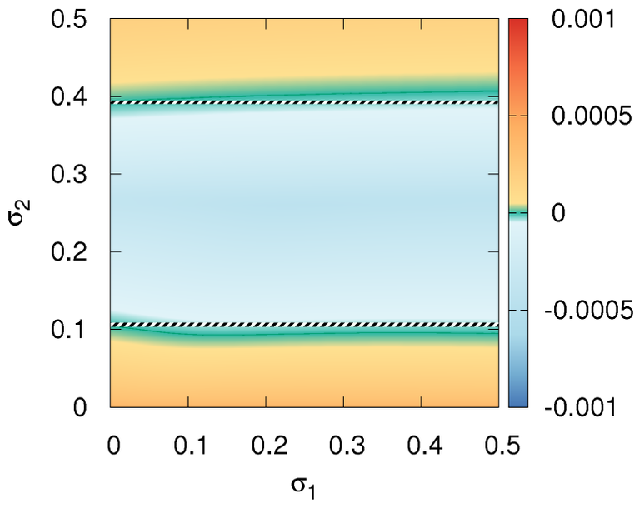}
\caption{Maximum Lyapunov exponent for ER-ER topologies
in Case~1 (top panel) and Case~2 (bottom panel). The darker blue
lines mark the points in the $(\sigma_1,\sigma_2)$ space
where $\Lambda$ vanishes, while the striped lines indicate
the critical values of $\sigma_2$ if layer~2 is considered
in isolation (or, equivalently, if $\sigma_1=0$).}\label{case1-2}
\end{figure}

\begin{figure}[t]
\centering
\includegraphics[width=0.35\textwidth]{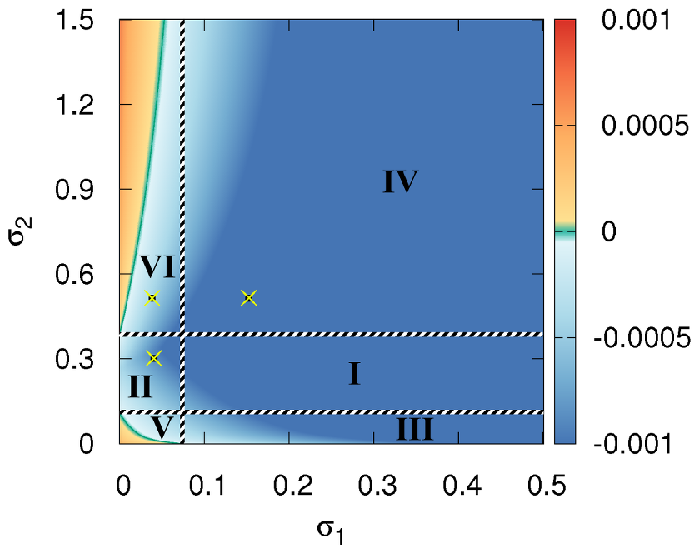}
\includegraphics[width=0.35\textwidth]{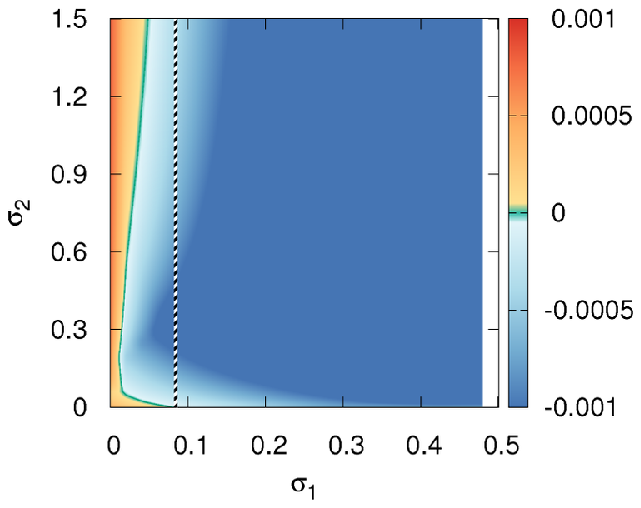}
\caption{Maximum Lyapunov exponent in Case~3 for ER-ER
and SF-SF topologies (top and bottom panel, respectively).
The darker blue lines mark the points in the $(\sigma_1,\sigma_2)$
plane where the maximum Lyapunov exponent is~0, while
the striped lines indicate the stability limits for the
$\sigma_1=0$ and $\sigma_2=0$. The points marked in the
top panel indicate the choices of coupling strengths used
for the numerical validation of the model. Note that for
SF networks in class III, the stability window disappears.}\label{case3}
\end{figure}
\begin{figure*}[t]
\centering
\includegraphics[width=0.9\textwidth]{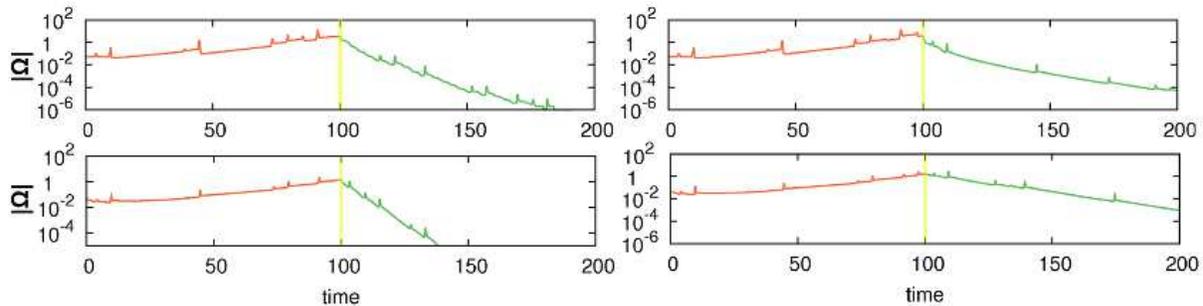}
\caption{Numerical validation of the stability
analysis. The error of synchronization increases
as long as the only active layer is the one predicted
to be unstable. When the other layer is switched
on, at time~100, the error of synchronization
decays exponentially towards~0, as predicted by
the model. With respect to Fig.~\ref{case3}, the
top-left panel corresponds to region~II, where
layer~1 is unstable and layer~2 stable, and the
interaction strengths used were $\sigma_1=0.04$
and $\sigma_2=0.3$. The bottom-left panel corresponds
to region~IV, where layer~1 is stable and layer~2
is unstable, and the interaction strengths were
$\sigma_1=0.15$ and $\sigma_2=0.5$. The top-right
and bottom-right panels correspond to region~VI,
where both layers are unstable. The layer active
from the beginning was layer~1 for the top-right
panel and layer~2 for the bottom-right. In both
cases the interaction strengths were $\sigma_1=0.04$
and $\sigma_2=0.5$.}\label{valid}
\end{figure*}
\textbf{Case 1}. Rewriting the system of equations~(\ref{mainsystem})
explicitly for each component of the $\boldsymbol\eta_{j}$,
we obtain here:
\begin{align}
 {{}\dot\eta_j}_1 &= -{\eta_j}_2-{\eta_j}_3\;,\label{eqs11}\\
 {{}\dot\eta_j}_2 &= {\eta_j}_1+0.2{\eta_j}_2-\sigma_2\sum_{k=2}^N\sum_{r=2}^N\lambda_r^{(2)}\Gamma_{r,k}\Gamma_{r,j}{\eta_k}_2\;,\label{eqs12}\\
 {{}\dot\eta_j}_3 &= s_3{\eta_j}_1+\left(s_1-9\right){\eta_j}_3-\sigma_1\lambda_j^{(1)}{\eta_j}_3\label{eqs13}\:,
\end{align}
from which the maximum Lyapunov exponent can be numerically
calculated.
In the top panel of Fig.~\ref{case1-2} we observe that, for ER-ER topologies, the first
layer is dominated by the second, as the stability region of
the whole system appears to be almost independent of $\sigma_1$,
disregarding a slight increase of the critical value of
$\sigma_2$ as $\sigma_1$ increases. This demonstrates the ability
of class~II systems to control the instabilities inherent to systems in class~I. This result appears
to be robust with respect to the choice of underlying structures,
as qualitatively similar results are obtained for SF-SF, ER-SF
and SF-ER topologies (see Fig.~1 in Supplementary Material).

\textbf{Case 2}. For Case~2, the system of equations~(\ref{mainsystem}) read:
\begin{align}
 {{}\dot\eta_j}_1 &= -{\eta_j}_2-{\eta_j}_3-\sigma_2\sum_{k=2}^N\sum_{r=2}^N\lambda_r^{(2)}\Gamma_{r,k}\Gamma_{r,j}{\eta_k}_1\label{eqs21}\;,\\
 {{}\dot\eta_j}_2 &= {\eta_j}_1+0.2{\eta_j}_2\label{eqs22}\;,\\
 {{}\dot\eta_j}_3 &= s_3{\eta_j}_1+\left(s_1-9\right){\eta_j}_3-\sigma_1\lambda_j^{(1)}{\eta_j}_3\label{eqs23}\:.
\end{align}
From the bottom panel in Fig.~\ref{case1-2} we observe that,
also in this case, the second layer strongly dominates
the whole system, as the overall stability window is
almost independent from the value of $\sigma_1$. This result,
together with that obtained for Case~1, suggests that class~I systems,
even though intrinsically preventing synchronization,
are easily controllable by both class~II and class~III
systems, even though, in analogy to the Case~1,
we observe a slight widening of the stability window
for increasing values of $\sigma_1$. Again, the results
are almost independent from the choice of the underlying topologies (see
Fig.~2 in the Supplementary Material).

\textbf{Case 3}. Finally, for Case~3, equations~(\ref{mainsystem}) become:
\begin{align}
 {{}\dot\eta_j}_1 &= -{\eta_j}_2-{\eta_j}_3-\sigma_2\sum_{k=2}^N\sum_{r=2}^N\lambda_r^{(2)}\Gamma_{r,k}\Gamma_{r,j}{\eta_k}_1\label{eqs31}\\
 {{}\dot\eta_j}_2 &= {\eta_j}_1+0.2{\eta_j}_2-\sigma_1\lambda_j^{(1)}{\eta_j}_2\label{eqs32}\\
 {{}\dot\eta_j}_3 &= s_3{\eta_j}_1+\left(s_1-9\right){\eta_j}_3\label{eqs33}\:.
\end{align}
Here, the system reveals its most striking features.
In particular, for ER-ER topologies (see Fig.~\ref{case3},
top panel), we observe~6 different regions, identified
in the figure by Roman numerals. Namely, in region~I, synchronization
is stable in both layers taken individually (or, equivalently,
for either $\sigma_1=0$ and $\sigma_2=0$), and, not
surprisingly, the full bi-layered network is also stable. Regions~II,
III and~IV correspond to scenarios qualitatively similar
to the ones seen previously, i.e.,  where stability properties
of one layer dominate over those of the other. Finally, regions~V and~VI
are the most important, as within them one finds effects that are genuinely due to the
multi-layered nature of the interactions. There, both layers are individually
unstable, and synchronization would not be observed at all for either
$\sigma_1=0$ or $\sigma_2=0$. However, the emergence
of a collective synchronous motion is remarkably obtained
with a suitable tuning of the parameters. In these
regions, it is therefore the \emph{simultaneous} action
of the two layers that induces stability.

Taken collectively, the results we obtained for
the three cases indicate that a multi-layer interaction
topology enhances the stability of the synchronized
state, even allowing the possibility of stabilizing
systems that are unstable when considered isolated.

\subsection{Numerical validation}
We validate the stability predictions derived from equations~(\ref{mainsystem}) by simulating
the full non-linear system of equations~(\ref{eomsys}) for an ER-ER topology in Case~3, with three
different choices of coupling constants $\sigma_1$ and $\sigma_2$. The three specific sets of
coupling values (shown in the top panel of Fig.~\ref{case3}) correspond to situations in which
either one or both layers are unstable when isolated, but yield a stable synchronized state when coupled.
More specifically, we have chosen ($\sigma_1=0.04$, $\sigma_2=0.3$)
corresponding to region~II, ($\sigma_1=0.15$, $\sigma_2=0.5$) in region~IV, and  ($\sigma_1=0.04$,
$\sigma_2=0.5$) in region~VI.

For all the three cases we run the simulations initially with the presence of only the unstable layer,
by setting either $\sigma_1=0$ or $\sigma_2=0$ depending on the set of couplings considered. Let us note
that for the third set of couplings (region~VI) either layer can be the initially active one, since both
are unstable when isolated. Then, after~100 integration steps, we activate the other layer by setting
its interaction strength to the (non-zero) value corresponding to the region for which we predicted a
stable synchronized state. As the systems evolve, we monitor the evolution of the norm $\left|\boldsymbol\Omega\right|\left(t\right)$
to evaluate the deviation from the synchronized solution with time.

\begin{figure}[t]
\centering
\includegraphics[width=0.45\textwidth]{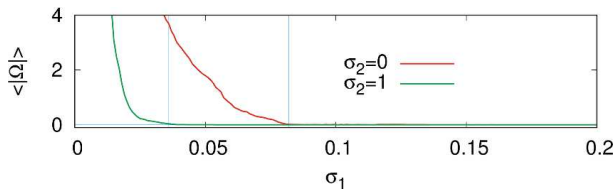}
\caption{Identification of the critical points.
For a system with ER-ER topology in Case~3 and
fixed $\sigma_2=1$, the synchronization error never
vanishes if $\sigma_1<\sigma_C\approx 0.04$. Conversely,
as soon as $\sigma_1>\sigma_C$, the system is again
able to synchronize (green line). One recovers
the mono-layer case by imposing $\sigma_2=0$, for
which similar results are found, with a critical
coupling strength of approximately~$0.08$ (red line). Both
results are in perfect agreement with the theoretical
predictions (see Fig.~\ref{valid}).}
\label{scan}
\end{figure}
The results, in Fig.~\ref{valid}, show that, when only the unstable interaction layer is active,
$\left|\boldsymbol\Omega\right|\left(t\right)$ never vanishes. However, as soon
as the other layer is switched on, the norm of $\boldsymbol\Omega$ undergoes a sudden change
of behaviour, starting an exponential decay towards~0. This confirms the prediction that
the unstable behaviour induced by each layer is compensated by the mutual presence of two
interaction layers.

Qualitatively similar scenarios are observed in Case~3 for SF-SF topologies,
as well as for ER-SF and SF-ER structures (see Fig.~3 in Supplementary Material).
Again, they confirm the correctness of the predictions, showing that in region~I
layer~1 dominates over layer~2, and that in region~II the overall stability can
be induced even when both layers are unstable in isolation.

To provide an even stronger demonstration of the predictive
power of our method, we simulate the full system for the ER-ER topology
in Case~3 fixing the value of $\sigma_2$ to~1 and varying the value of
$\sigma_1$ from~0 to~$0.2$. Starting from an initial perturbed synchronized
state, after a transient of~100 time units we measure the average of $\left|\boldsymbol\Omega\right|$
over the next~20 integration steps. The results, in Fig.~\ref{scan}, show
a very good agreement between the simulations and the theoretical predicion
(cf.\ Fig.~\ref{valid}).
For values of $\sigma_1$ less then a critical value of approximately~$0.04$,
the system never synchronizes. Conversely, when $\sigma_1$ crosses the critical
value, the system is able to reach a synchronized state. Interestingly, repeating the
simulation with $\sigma_2=0$ one recovers the monoplex case. Also in this instance,
we find good agreement between theoretical prediction and simulation, with a
critical coupling value of approximately~$0.08$.

\section{Discussion}
The results shown above clearly illustrate the rich dynamical phenomenology
that emerges when the multi-layer structure of real networked systems is taken
into account. In an explicit example, we have observed that synchronization
stability can be induced in unstable networked layers by coupling them with
stable ones. In addition, we have shown that stability can be achieved even
when all the layers of a complex system are unstable if considered in isolation.
This latter result constitutes a clear instance of an effect that is intrinsic
to the true multi-layer nature of the interactions amongst the dynamical units.
Similarly, we expect that the opposite could also be observed,
namely that the synchronizability of a system decreases, or even disappears,
when two individually synchronizable layers are combined.

On more general grounds, the theory developed here allows one to assess
the stability of the synchronized state of coupled non-linear dynamical
systems with multi-layer interactions in a fully general setting. The system
can have any arbitrary number of layers and, perhaps more importantly,
the network structures of each layer can be fully independent, as we do
not exploit any special structural or dynamical property to develop our
theory. This way, our approach generalizes the celebrated Master Stability
Function~\cite{PeC998} to multi-layer structures, retaining the general
applicability of the original method. The complexity in the extra layers
is reflected in the fact that the formalism yields a set of coupled linear
differential equations (Eq.~\ref{mainsystem}), rather than a single parametric
variational equation, which is recovered in the case of commuting
Laplacians.
This system of equations describes the evolution of a set of $d$-dimensional
vectors that encode the displacement of each dynamical system from the
synchronized state. The solution of the system gives a
necessary condition for stability: if the norm of these vectors vanishes in time, then the system
gets progressively closer to synchronization, which is therefore stable;
if, instead, the length of the vectors always remains greater than~0, then
the synchronized state is unstable.

The generality of the method presented, which is applicable
to any undirected structure, and its straightforward implementation
for any choice of $C^1$ dynamical setup pave the way for the exploration of synchronization
properties on multi-layer networks of arbitrary size and structure.
Thus, we are confident that our work can be used in the
design of optimal multilayered synchronizable systems, a problem that has attracted
much attention in mono-layer complex networks~\cite{design1,design2,design3,design4}.
In fact, the straightforward nature of our formalism makes it suitable to be efficiently
used together with successful techniques, such as the rewiring of links or the search for an optimal distribution
of links weights, in the context of multilayer networks.
In turn, these techniques may help in addressing the already-mentioned question
of the suppression of synchronization due to the interaction between layers,
unveiling possible combinations of stable layers that, when interacting, suppress
the dynamical coherence that they show in isolation.
Also, we believe that the reliability of our method will
provide aid to the highly current field of multiplex network controllability~\cite{LSB011,Nepusz012,Sun013,Gao014,Skardal015},
enabling researchers to engineer control layers to drive the system dynamics towards a desired state.

In addition, several extensions of our work towards more general systems are possible.
A particularly relevant one is the study of multi-layer networks of heterogeneous oscillators, which
have a rich phenomenology, and whose synchronizability has been shown
to depend on all the Laplacian eigenvalues~\cite{Ska14}, in a way similar to
the results presented here.
Relaxing the requirement of an undirected structure,
our approach can also be used to study directed networks.
The graph Laplacians in this case are not necessarily
diagonalizable, but a considerable amount of information
can be still extracted from them using singular value decomposition.
For example, it is already known that directed networks
can be rewired to obtain an optimal distribution of
in-degrees for synchronization~\cite{Ska16}. Further
areas that we intend to explore in future work are
those of almost identical oscillators and almost identical
layers, which can be approached using perturbative methods
and constitute more research directions with even wider applicability.

Finally, as our method allows one to study the rich synchronization
phenomenology of general multi-layer networks, we believe it will find application
in technological, biological and social systems where synchronization processes and
multilayered interactions are at work. Some examples are coupled power-grid and communication
systems, some brain neuropathologies such as epilepsy,
and the onset of coordinated social behavior when multiple interaction
channels coexist. Of course, as mentioned above, these applications will demand further
advances in order to include specific features such as the non-homogeneity of interacting
units or the possibility of directional interactions.

\section{Materials and Methods}
\subsection{Linearization around the synchronized solution}
To linearize the system in Eq.~(\ref{mainsystem}) around the synchronization
manifold, use the fact that for any $C^1$-vector field $\mathbf f$ we can write:
\begin{equation*}
 \mathbf f\left(\mathbf x\right)\approx\mathbf f\left(\mathbf{x_0}\right)+J\mathbf f\left(\mathbf{x_0}\right)\cdot\left(\mathbf x-\mathbf{x_0}\right)\:.
\end{equation*}
Using this relation, we can expand $\mathbf F$ and $\mathbf H$
around $\mathbf s$ in the system of equations~\ref{mainsystem}
to obtain:
\begin{equation}
 \delta\dot{\mathbf x}_i = \dot{\mathbf x}_i-\dot{\mathbf s}\approx J\mathbf F\left(\mathbf s\right)\cdot\delta\mathbf{x}_i-\sum_{\alpha=1}^M\sigma_\alpha J\mathbf H_{\alpha}\left(\mathbf s\right)\cdot\sum_{j=1}^N L^{(\alpha)}_{i,j}\delta\mathbf{x}_j\:.
\end{equation}
Now, we use the Kronecker matrix product to decompose
the equation above into self-mixing and interaction
terms, and introduce the vector $\delta\mathbf X$,
to get the final system of equations~\ref{linglob}.

The system~\ref{linglob} can be rewritten by projecting
$\delta\mathbf X$ onto the Laplacian eigenvectors
of a layer. The choice of layer to carry out this projection
is entirely arbitrary, because the Laplacian eigenvectors are
always a basis of $\mathbb{R}^N$. Without loss of generality,
we choose here layer~1, and we ensure that the eigenvectors are
orthonormal. Then, define $\mathds 1_d$ to be the $d$-dimensional
identity matrix, and multiply Eq.~\ref{linglob} on the
left by $\left({\mathbf{V}^{(1)}}^\mathrm{T}\otimes\mathds 1_d\right)$:
\begin{multline*}
 \left({\mathbf{V}^{(1)}}^\mathrm{T}\otimes\mathds 1_d\right)\delta\mathbf{\dot X} = \Bigg[\left({\mathbf{V}^{(1)}}^\mathrm{T}\otimes\mathds 1_d\right)\left(\mathds 1\otimes J\mathbf F\left(\mathbf s\right)\right)\\
 \left. -\sum_{\alpha=1}^M\sigma_\alpha\left({\mathbf{V}^{(1)}}^\mathrm{T}\otimes\mathds 1_d\right)\left(\mathbf L^{(\alpha)}\otimes J\mathbf H_\alpha\left(\mathbf s\right)\right)\right]\delta\mathbf X\:.
\end{multline*}
Now, use the relation
\begin{equation}\label{kronmurb}
 \left(\mathbf{M_1}\otimes\mathbf{M_2}\right)\left(\mathbf{M_3}\otimes\mathbf{M_4}\right) = \left(\mathbf{M_1}\mathbf{M_3}\right)\otimes\left(\mathbf{M_2}\mathbf{M_4}\right)
\end{equation}
to obtain
\begin{multline*}
 \left({\mathbf{V}^{(1)}}^\mathrm{T}\otimes\mathds 1_d\right)\delta\mathbf{\dot X} = \left[{\mathbf{V}^{(1)}}^\mathrm{T}\otimes J\mathbf F\left(\mathbf s\right)\right.\\
 \left. -\left(\sigma_1\mathbf{D}^{(1)}{\mathbf{V}^{(1)}}^\mathrm{T}\right)\otimes J\mathbf{H}_1\left(\mathbf s\right)\right]\delta\mathbf X\\
 -\sum_{\alpha=2}^M\sigma_\alpha\left({\mathbf{V}^{(1)}}^\mathrm{T}\mathbf L^{(\alpha)}\right)\otimes J\mathbf H_\alpha\left(\mathbf s\right)\delta\mathbf X\:,
\end{multline*}
where $\mathbf{D}^{(\alpha)}$ is the diagonal matrix
of the eigenvalues of layer $\alpha$, and we have split
the sum into the first term and the remaining $M-1$ terms.
Left-multiply
the first occurrence of ${\mathbf{V}^{(1)}}^\mathrm{T}$
in the right-hand-side by $\mathds 1$, and right-multiply
$\mathbf F$ and $\mathbf H_1$ by $\mathds 1_d$. Then,
using again Eq.~\ref{kronmurb}, it is
\begin{multline*}
 \left({\mathbf{V}^{(1)}}^\mathrm{T}\otimes\mathds 1_d\right)\delta\mathbf{\dot X} = \left[\left(\mathds 1\otimes J\mathbf F\left(\mathbf s\right)\right)\left({\mathbf{V}^{(1)}}^\mathrm{T}\otimes\mathds 1_d\right)\right.\\
 \left. -\left(\sigma_1\mathbf{D}^{(1)}\otimes J\mathbf H_1\left(\mathbf s\right)\right)\left({\mathbf{V}^{(1)}}^\mathrm{T}\otimes\mathds 1_d\right)\right]\delta\mathbf X\\
 -\sum_{\alpha=2}^M\sigma_\alpha{\mathbf{V}^{(1)}}^\mathrm{T}\mathbf L^{(\alpha)}\otimes J\mathbf H_\alpha\left(\mathbf s\right)\delta\mathbf X\:.
\end{multline*}
Factor out $\left({\mathbf{V}^{(1)}}^\mathrm{T}\otimes\mathds 1_d\right)$ to get
\begin{multline*}
 \left({\mathbf{V}^{(1)}}^\mathrm{T}\otimes\mathds 1_d\right)\delta\mathbf{\dot X} = \left(\mathds 1\otimes J\mathbf F\left(\mathbf s\right)-\sigma_1\mathbf D^{(1)}\otimes J\mathbf H_1\left(\mathbf s\right)\right)\\
 \times\left({\mathbf{V}^{(1)}}^\mathrm{T}\otimes\mathds 1_d\right)\delta\mathbf X -\sum_{\alpha=2}^M\sigma_\alpha{\mathbf{V}^{(1)}}^\mathrm{T}\mathbf L^{(\alpha)}\otimes J\mathbf H_\alpha\left(\mathbf s\right)\delta\mathbf X\:.
\end{multline*}
The relation
\begin{equation*}
 \left(\mathbf{M_1}\otimes\mathbf{M_2}\right)^{-1} = \mathbf{M_1}^{-1}\otimes\mathbf{M_2}^{-1}
\end{equation*}
implies that $\left({\mathbf{V}^{(1)}}\otimes\mathds 1_d\right)\left({\mathbf{V}^{(1)}}^\mathrm{T}\otimes\mathds 1_d\right)$
is the $mN$-dimensional identity matrix. Then, left-multiply the last
last $\delta\mathbf X$ by this expression, obtaining
\begin{multline*}
 \left({\mathbf{V}^{(1)}}^\mathrm{T}\otimes\mathds 1_d\right)\delta\mathbf{\dot X} = \left(\mathds 1\otimes J\mathbf F\left(\mathbf s\right)-\sigma_1\mathbf D^{(1)}\otimes J\mathbf H_1\left(\mathbf s\right)\right)\\
 \times\left({\mathbf{V}^{(1)}}^\mathrm{T}\otimes\mathds 1_d\right)\delta\mathbf X -\sum_{\alpha=2}^M\sigma_\alpha{\mathbf{V}^{(1)}}^\mathrm{T}\mathbf L^{(\alpha)}\otimes J\mathbf H_\alpha\left(\mathbf s\right)\\
 \times\left({\mathbf{V}^{(1)}}\otimes\mathds 1_d\right)\left({\mathbf{V}^{(1)}}^\mathrm{T}\otimes\mathds 1_d\right)\delta\mathbf X\:.
\end{multline*}

Now define the vector-of-vectors
\begin{equation*}
 \boldsymbol\eta\equiv\left({\mathbf{V}^{(1)}}^\mathrm{T}\otimes\mathds 1_d\right)\delta\mathbf X\:.
\end{equation*}
Each component of $\boldsymbol\eta$ is the projection
of the global synchronization error vector $\delta\mathbf X$
onto the space spanned by the corresponding Laplacian
eigenvector of layer~1. The first eigenvector, which
defines the synchronization manifold, is common to all
layers, and all other eigenvectors are orthogonal to
it. Thus, the norm of the projection of $\boldsymbol\eta$ over the
space spanned by the last $N-1$ eigenvectors is a measure
of the synchronization error in the directions transverse
to the synchronization manifold. Because of how $\eta$
is built, this projection is just the vector $\boldsymbol\Omega$,
consisting of the last $N-1$ components of $\boldsymbol\eta$.
With this definition of $\boldsymbol\eta$,
left-multiply $\mathbf L^{(\alpha)}$ by
the identity expressed as $\mathbf V^{(\alpha)}{\mathbf V^{(\alpha)}}^\mathrm{T}$,
to obtain
\begin{multline*}
 \dot{\boldsymbol\eta} = \left(\mathds 1\otimes J\mathbf F\left(\mathbf s\right)-\sigma_1\mathbf D^{(1)}\otimes J\mathbf H_1\left(\mathbf s\right)\right)\boldsymbol\eta\\
 -\sum_{\alpha=2}^M\sigma_\alpha{\mathbf V^{(1)}}^\mathrm{T}{\mathbf V^{(\alpha)}}\mathbf D^{(\alpha)}{\mathbf V^{(\alpha)}}^\mathrm{T}{\mathbf V^{(1)}}\otimes J\mathbf H_\alpha\left(\mathbf s\right)\boldsymbol\eta\:.
\end{multline*}
In this vector equation, the first part is purely variational,
since it consists of a block-diagonal matrix that multiplies
the vector-of-vectors $\boldsymbol\eta$. The second part,
instead, mixes different components of $\boldsymbol\eta$.
This can be seen more easily expressing the vector equation as
a system of equations, one for each component $j$ of $\boldsymbol\eta$.

To write such a system, it is convenient to first define
$\boldsymbol\Gamma^{(\alpha)}\equiv{\mathbf V^{(\alpha)}}^\mathrm{T}{\mathbf V^{(1)}}$.
Then, consider the non-variational part.
Its contribution to $j$th component of $\dot{\boldsymbol\eta}$
is given by the product of the $j$th row of blocks of the
block-matrix by $\boldsymbol\eta$. In turn, each element
of this row of blocks consists of the corresponding element of
the $j$th row of ${\boldsymbol\Gamma^{(\alpha)}}^\mathrm{T}\mathbf D^{(\alpha)}\boldsymbol\Gamma^{(\alpha)}$
multiplied by $J\mathbf H_\alpha\left(\mathbf s\right)$:
\begin{equation*}
 \left({\boldsymbol\Gamma^{(\alpha)}}^\mathrm{T}\mathbf D^{(\alpha)}\boldsymbol\Gamma^{(\alpha)}\right)_{j,k} = \sum_{r=1}^N{\Gamma^{(\alpha)}}^\mathrm{T}_{j,r}\lambda_r^{(\alpha)}\Gamma^{(\alpha)}_{r,k}\:.
\end{equation*}
Summing over all the components $\boldsymbol\eta_k$ yields
\begin{multline*}
 \dot{\boldsymbol\eta}_{j} = \left(J\mathbf F\left(\mathbf s\right)-\sigma_1\lambda_j^{(1)}J\mathbf{H}_1\left(\mathbf s\right)\right)\boldsymbol\eta_j+\\
 -\sum_{\alpha=2}^M\sigma_\alpha\sum_{k=2}^N\sum_{r=2}^N\lambda_r^{(\alpha)}\Gamma^{(\alpha)}_{r,k}\Gamma^{(\alpha)}_{r,j}J\mathbf H_{\boldsymbol\alpha}\left(\mathbf s\right)\boldsymbol\eta_k\:,
\end{multline*}
which is Eq.~\ref{mainsystem}. Notice that the sums over
$r$ and $k$ start from~2, because the first eigenvalue is
always~0, and the orthonormality of the eigenvectors guarantees
that all the elements of the first column of $\boldsymbol\Gamma^{(\alpha)}$
except the first are~0. Each matrix $\Gamma^{(\alpha)}$ effectively captures the alignment
of the Laplacian eigenvectors of layer~$\alpha$ with those
of layer~1. If the eigenvectors for layer~$\alpha$ are identical
to those of layer~1, as it happens when the two Laplacians commute,
then $\Gamma^{(\alpha)}=\mathds 1$. Of course, one can generalize
the definition of $\Gamma^{(\alpha)}$ to consider any two layers,
introducing the matrices
$\Xi^{(\alpha,\beta)}\equiv{\mathbf V^{(\alpha)}}^\mathrm{T}{\mathbf V^{(\beta)}}=\Gamma^{(\alpha)}{\Gamma^{(\beta)}}^\mathrm{T}$
that can be even used to define a measure $\ell_D$ of ``dynamical distance'' between
two layers $\alpha$ and $\beta$:
\begin{equation*}
 \ell_D = \sum_{i=2}^N\left[\sum_{j=2}^N \left(\Xi^{(\alpha,\beta)}_{i,j}\right)^2\right]-\left(\Xi^{(\alpha,\beta)}_{i,i}\right)^2\:.
\end{equation*}

\subsection{MSF and stability classes}
A particular case of the treatment we considered
above happens when $M=1$. In this case, the second
term on the right-hand side of Eq.~\ref{mainsystem}
disappears, and the system takes the variational
form $\dot{\boldsymbol\eta}_{i}=\mathbf{K}_i\boldsymbol\eta_{i}$,
where $\mathbf{K}_i\equiv J\mathbf F\left(\mathbf s\right)-\sigma\lambda_iJ\mathbf H\left(\mathbf s\right)$
is an evolution kernel evaluated on the synchronization
manifold. Since $\lambda_1=0$, this equation separates
the contribution parallel to the manifold, which
reduces to $\dot{\boldsymbol\eta}_{1}=J\mathbf F\left(\mathbf s\right)\boldsymbol\eta_{1}$,
from the other $N-1$, which describe perturbations
in the directions transverse to the manifold, and
that have to be damped for the synchronized state
to be stable. Since the Jacobians of $\mathbf F$
and $\mathbf H$ are evaluated on the synchronized
state, the variational equations differ only in
the eigenvalues $\lambda_i$. Thus, one can extract
from each of them a set of $d$ conditional Lyapunov
exponents, evaluated along the eigen-modes associated
to $\lambda_i$. Putting $\nu\equiv\sigma\lambda_i$,
the parametrical behaviour of the largest of these
exponents $\Lambda\left(\nu\right)$ defines the
so-called Master Stability Function (MSF)~\cite{PeC998}.
If the network is undirected, then the spectrum
of the Laplacian is real, and the MSF is a real
function of $\nu$. Crucially, for all possible choices of
$\mathbf F$ and $\mathbf H$, the MSF of a network falls into one of three possible behaviour classes,
defined as follows~\cite{Boc006}:
\begin{itemize}
 \item Class~I: $\Lambda\left(\nu\right)$ never intercepts the $x$ axis.
 \item Class~II: $\Lambda\left(\nu\right)$ intercepts the $x$ axis in a single point at some $\nu_c \geqslant 0$.
 \item Class~III: $\Lambda\left(\nu\right)$ is a convex function with negative values
 within some window $\nu_{c1}<\nu<\nu_{c2}$; in general, $\nu_{c1}\geqslant 0$,
 with the equality holding when $\mathbf F$ supports a periodic motion.
\end{itemize}
The elegance of the MSF formalism manifests itself
at its finest for systems in Class~III, for which
synchronization is stable only if $\sigma\lambda_2>\nu_{c1}$
and $\sigma\lambda_N<\nu_{c2}$ hold simultaneously.
This condition implies ${}^{\lambda_N}/{}_{\lambda_2}<{}^{\nu_{c2}}/{}_{\nu_{c1}}$.
Since ${}^{\lambda_N}/{}_{\lambda_2}$ is entirely
determined by the network topology and ${}^{\nu_{c2}}/{}_{\nu_{c1}}$
depends only on the dynamical functions $\mathbf F$
and $\mathbf H$, one has a simple stability criterion
in which structure and dynamics are decoupled.

\subsection{Network generation}
To generate the networks for our simulations,
we use the algorithm described in Ref.~\cite{Gom006},
that creates a one-parameter family of complex
networks with a tunable degree of heterogeneity.
The algorithm works as follows: start from a
fully connected network with $m_0$ nodes, and
a set $\mathcal X$ containing $N-m_0$ isolated
nodes. At each time step, select a new node from
$\mathcal X$, and link it to $m$ other nodes,
selected amongst all other nodes. The choice
of the target nodes happens uniformly at random
with probability $\alpha$, and following a preferential
attachment rule with probability $1-\alpha$.
Repeating these steps $N-m_0$ times, one obtains
networks with the same number of nodes and links,
whose structure interpolates between ER, for
$\alpha=1$, and SF, for $\alpha=0$.

\subsection{Numerical calculations}
To compute the maximum Lyapunov exponent for a given
pair of coupling strengths $\sigma_1$ and $\sigma_2$,
we first integrate a single Rössler oscillator from
an initial state $\left(0,0,0\right)$ for a transient
time $t_\mathrm{trans}$, sufficient to reach the chaotic attractor.
The integration is carried out using a fourth-order
Runge-Kutta integrator with a time step of $5\times 10^{-3}$,
for which we choose a transient time $t_\mathrm{trans}=300$.
Then, we integrate the systems for the perturbations
(Eqs.~\ref{eqs11}--\ref{eqs13}, \ref{eqs21}--\ref{eqs23}
and~\ref{eqs31}--\ref{eqs33}) using Euler's method,
again with a same time-step of $5\times 10^{-3}$. The
initial conditions are so that all the components of
all the $\boldsymbol\eta_{j}$ are $1/\sqrt{3\left(N-1\right)}$,
making $\boldsymbol\Omega$ a unit vector. At the same
time, we continue the integration of the single Rössler
unit, to provide for $s_1$ and $s_3$, that appear in
the perturbation equations. This process is repeated
for~500 time windows, each of the duration of~1 unit
(200~steps). After each window $n$ we compute the norm
of the overall perturbation $\left|\boldsymbol\Omega\right|\left(n\right)$,
and re-scale the components of the $\boldsymbol\eta_{j}$
so that at the start of the next time window the norm
of $\boldsymbol\Omega$ is again set to~1. Finally, when
the integration is completed, we estimate the maximum
Lyapunov exponent as
\begin{equation*}
 \Lambda = \frac{1}{500}\sum_{n=1}^{500}\log\left(\left|\boldsymbol\Omega\right|\left(n\right)\right)\:.
\end{equation*}

\section*{Acknowledgements}
The authors would like to express their gratitude to Alex Arenas
and Javier Buldú for many interesting and fruitful discussions.

\section*{Funding}
The work of JGG was supported by the Spanish MINECO via grants
FIS2012-38266-C02-01 and FIS2011-25167, and by the European AQ38
Union through FET Proactive Project MULTIPLEX (Multilevel Complex Networks and Systems),
contract no.~317532.

\section*{Author contribution}
CIDG and SB developed the theory. SB designed the simulations.
JGG implemented and carried out the simulations, and analyzed
the results. All authors wrote the manuscript.

\section*{Competing interests}
The authors declare they have no competing interests.

\section*{Data and material availability}
All data are present in the paper and in the supplementary material.

\begin{thebibliography}{99}
\bibitem{Str001} S.~H. Strogatz. Exploring complex networks. \textit{Nature} \textbf{410}, 268--276 (2001).
\bibitem{AlB002} R. Albert, A.-L. Barabási. Statistical mechanics of complex networks. \textit{Rev.\ Mod.\ Phys.} \textbf{74}, 47--97 (2002).
\bibitem{New003} M.~E.~J. Newman. The structure and function of complex networks. \textit{SIAM Rev.} \textbf{45}, 167--256 (2003).
\bibitem{DoM003} S.~N. Dorogovtsev, J.~F.~F. Mendes. \textit{Evolution of Networks: From Biological Nets to the Internet and WWW}. (Oxford University Press, Oxford, UK, 2003).
\bibitem{Ben004} E. Ben-Naim, H. Frauenfelder, Z. Toroczkai. \textit{Complex Networks} (Lecture Notes in Physics, Vol.~650) (Springer, Berlin, Germany) (2004).
\bibitem{Boc006} S.~Boccaletti, V.~Latora, Y.~Moreno, M.~Chavez, D.~U. Hwang. Complex networks: Structure and dynamics. \textit{Phys.\ Rep.} \textbf{424}, 175--308 (2006).
\bibitem{Cal007} G. Caldarelli. \textit{Scale-free Networks: Complex Webs in Nature and Technology}. (Cambridge University Press, Cambridge, UK, 2007).
\bibitem{New010} M.~E.~J. Newman. \textit{Networks: An Introduction.} (Oxford University Press, New York, 2010).
\bibitem{CoH010} R. Cohen, S. Havlin. \textit{Complex Networks: Structure, Robustness and Function}. (Cambridge University Press, Cambridge, UK, 2010).
\bibitem{WaS998} D. Watts, S.~H. Strogatz. Collective dynamics of ``small-world'' networks. \textit{Nature} \textbf{393}, 440--442 (1998).
\bibitem{BaA999} A.-L. Barabási, R. Albert. Emergence of scaling in random networks. \textit{Science} \textbf{286}, 509--512 (1999).
\bibitem{DGM008} S.~N. Dorogovtsev, A.~V. Goltsev, J.~F.~F. Mendes. Critical phenomena in complex networks. \textit{Rev.\ Mod.\ Phys.} \textbf{80}, 1275--1335 (2008).
\bibitem{Gui005} R. Guimerá, L.~A.~N. Amaral. Functional cartography of complex metabolic networks. \textit{Nature} \textbf{433}, 895--900 (2005).
\bibitem{For010} S. Fortunato. Community detection in graphs. \textit{Phys.\ Rep.} \textbf{486}, 75--174 (2010).
\bibitem{del13} C.~I. del~Genio, T. House. Endemic infections are always possible on regular networks. \textit{Phys.\ Rev.\ E} \textbf{88}, 040801 (2013).
\bibitem{Pei14} T.~P. Peixoto, Hierarchical block structures and high-resolution model selection in large networks. \textit{Phys. Rev. X} \textbf{4}, 011047 (2014).
\bibitem{Wil14} O. Williams, C.~I. del~Genio. Degree correlations in directed scale-free networks. \textit{PLoS One} \textbf{9}, e110121 (2014).
\bibitem{Tre15} S. Treviño~III, A. Nyberg, C.~I. del~Genio, K.~E. Bassler, Fast and accurate determination of modularity and its effect size. \textit{J. Stat.\ Mech.\ - Theory E.}, P02003 (2015).
\bibitem{New15} M.~E.~J. Newman, T.~P. Peixoto. Generalized Communities in Networks. \textit{Phys. Rev. Lett.} \textbf{115}, 088701 (2015).
\bibitem{BBV008} A. Barrat, M. Barthélemy, A. Vespignani, \textit{Dynamical Processes on Complex Networks.} (Cambridge University Press, Cambridge, UK, 2008).
\bibitem{BaB013} B. Barzel, A.-L. Barabási. Universality in network dynamics. \textit{Nat.\ Phys.} \textbf{9}, 673--681 (2013).
\bibitem{GBB016} J. Gao, B. Barzel, A.-L. Barabási. Universal resilience patterns in complex networks. \textit{Nature} \textbf{530}, 307--312 (2016).
\bibitem{CFL009} C. Castellano, S. Fortunato, V. Loreto. Statistical physics of social dynamics. \textit{Rev.\ Mod.\ Phys.} \textbf{81}, 591--646 (2009).
\bibitem{Pas015} R. Pastor-Satorras, C. Castellano, P. Van~Mieghem, A. Vespignani. Epidemic processes in complex networks. \textit{Rev.\ Mod.\ Phys.} \textbf{87}, 925--979 (2015).
\bibitem{Boc002} S. Boccaletti, J. Kurths, G. Osipov, D.L. Valladares, C.S. Zhou. The synchronization of chaotic systems. \textit{Phys.\ Rep.} \textbf{366}, 1--101 (2002).
\bibitem{LSB011} Y.-Y. Liu, J.-J. Slotine, A.-L. Barabási. Controllability of complex networks. \textit{Nature} \textbf{473}, 167--173 (2011).
\bibitem{Pik001} A. Pikovsky, M. Rosenblum, J. Kurths. \textit{Synchronization: A Universal Concept in Nonlinear Sciences}. (Cambridge University Press, Cambridge, UK, 2003).
\bibitem{Str003} S.~H. Strogatz. \textit{Sync: The Emerging Science of Spontaneous Order} (New York: Hyperion, 2003).
\bibitem{Man004} S.~C. Manrubia, A.S. Mikhailov, D.H. Zanette. \textit{Emergence of Dynamical Order. Synchronization Phenomena in Complex Systems} (World Scientific, Singapore, 2004)
\bibitem{PeC998} L.~M. Pecora, T.~L. Carroll. Master Stability Functions for Synchronized Coupled Systems. \textit{Phys.\ Rev.\ Lett.} \textbf{80}, 2109--2112 (1998).
\bibitem{Lag000} L.~F. Lago-Fernández, R. Huerta, F. Corbacho, J.~A. Sigüenza. Fast Response and Temporal Coherent Oscillations in Small-World Networks. \textit{Phys.\ Rev.\ Lett.} \textbf{84}, 2758--2761 (2000).
\bibitem{Bar002} M. Barahona, L.~M. Pecora. Synchronization in Small-World Systems. \textit{Phys.\ Rev.\ Lett.} \textbf{89}, 054101 (2002).
\bibitem{Nis003} T. Nishikawa, A.~E. Motter, Y.-C. Lai, F.~C. Hoppensteadt. Heterogeneity in Oscillator Networks: Are Smaller Worlds Easier to Synchronize? \textit{Phys.\ Rev.\ Lett.} \textbf{91}, 014101 (2003).
\bibitem{Bel004} I.~V. Belykh, V.~N. Belykh, M. Hasler. Blinking model and synchronization in small-world networks with a time-varying coupling. \textit{Physica~D} \textbf{195}, 188--206 (2004).
\bibitem{Hwa005} D.-U. Hwang, M. Chavez, A. Amann, and S. Boccaletti. Synchronization in Complex Networks with Age Ordering. \textit{Phys.\ Rev.\ Lett.} \textbf{94}, 138701 (2005).
\bibitem{Cha005} M. Chavez, D.-U. Hwang, A. Amann, H.~G.~E. Hentschel, S. Boccaletti. Synchronization is Enhanced in Weighted Complex Networks. \textit{Phys.\ Rev.\ Lett.} \textbf{94}, 218701 (2005).
\bibitem{Mot005} A.~E. Motter, C.-S. Zhou, J. Kurths. Enhancing complex-network synchronization. \textit{EPL} \textbf{69}, 334--337 (2005).
\bibitem{Zhou006} C. Zhou, A.E. Motter, J. Kurths. Universality in the Synchronization of Weighted Random Networks. \textit{Phys. Rev. Lett.} \textbf{96}, 034101 (2006).
\bibitem{Lod007} I. Lodato, S. Boccaletti, V. Latora. Synchronization Properties of Network Motifs. \textit{EPL} \textbf{78}, 28001 (2007).
\bibitem{JGG011} J. G\'omez-Garde\~nes, S. Gómez, A. Arenas, Y. Moreno. Explosive Synchronization Transitions in Scale-Free Networks. \textit{Phys. Rev. Lett.} \textbf{106}, 128701 (2011).
\bibitem{Bil14}  S. Bilal and R. Ramaswamy. Synchronization and amplitude death in hypernetworks. \textit{Phys. Rev.~E} \textbf{89}, 062923 (2014).
\bibitem{del015} C.~I. del~Genio, M. Romance, R. Criado, S. Boccaletti. Synchronization in dynamical networks with unconstrained structure switching. \textit{Phys.\ Rev.~E} \textbf{92}, 062819 (2015).
\bibitem{Boc014} S. Boccaletti, G. Bianconi, R. Criado, C.~I. del~Genio, J. Gómez-Gardeñes, M. Romance, I. Sendiña-Nadal, Z. Wang, M. Zanin. The structure and dynamics of multilayer networks. \textit{Phys.\ Rep.} \textbf{544}, 1--122 (2014).
\bibitem{Kiv014} M. Kivela, A. Arenas, M. Barthélemy, J.~P. Gleeson, Y. Moreno, M.~A. Porter. Multilayer networks. \textit{J. Complex Networks} \textbf{2}, 203--271 (2014).
\bibitem{Lee015} K.-M. Lee, B. Min, K.-I. Goh. Towards real-world complexity: an introduction to multiplex networks. \textit{Eur.\ Phys.\ J. B} \textbf{88}, 1--20 (2015).
\bibitem{Bia015} G. Bianconi. Interdisciplinary and physics challenges of network theory. \textit{EPL} \textbf{111}, 56001 (2015).
\bibitem{Sze010} M. Szell, R. Lambiotte, S. Thurner. Multirelational organization of large-scale social networks in an online world. \textit{Proc.\ Natl.\ Acad.\ Sci.\ USA} \textbf{107}, 13636--13641 (2010).
\bibitem{Cardillo} A. Cardillo, J. G\'omez-Garde\~nes, M. Zanin, M. Romance, D. Papo, F. del Pozo, S. Boccaletti
Emergence of network features from multiplexity. \textit{Sci. Rep.} \textbf{3}, 1344 (2013).
\bibitem{Hal014} A. Halu, S. Mukherjee, G. Bianconi. Emergence of overlap in ensembles of spatial multiplexes and statistical mechanics of spatial interacting network ensembles. \textit{Phys.\ Rev.~E} \textbf{89}, 012806 (2014).
\bibitem{Adh011} B.~M. Adhikari, A. Prasad, M. Dhamala. Time-delay-induced phase-transition to synchrony in coupled bursting neurons. \textit{Chaos} \textbf{21}, 023116 (2011).
\bibitem{Rad013} F. Radicchi, A. Arenas. Abrupt transition in the structural formation of interconnected networks. \textit{Nature Physics} \textbf{9}, 717 (2013).
\bibitem{JGG15} J. G\'omez-Garde\~nes, M. De Domenico, G. Guti\'errez, A. Arenas, S. G\'omez. Layer-layer competition in multiplex complex networks. \textit{Phil. Trans. R. Soc. A} \textbf{373}, 20150117 (2015).
\bibitem{Bul010} S.~V. Buldyrev, R. Parshani, G. Paul, H.~E. Stanley, S. Havlin. Catastrophic Cascade of Failures in Interdependent Networks. \textit{Nature} \textbf{464}, 1025--1028 (2010).
\bibitem{Son012} S.-W. Son, G. Bizhani, C. Christensen, P. Grassberger, M. Paczuski. Percolation Theory on Interdependent Networks Based on Epidemic Spreading. \textit{EPL} \textbf{97}, 16006 (2012).
\bibitem{Gao012} J. Gao, S.~V. Buldyrev, H.~E. Stanley, S. Havlin. Networks Formed from Interdependent Networks. \textit{Nat.\ Phys.} \textbf{8}, 40--48 (2012).
\bibitem{BiD014} G. Bianconi, S.~N. Dorogovtsev. Multiple percolation transitions in a configuration model of a network of networks. \textit{Phys.\ Rev.\ E}, \textbf{89}, 062814 (2014).
\bibitem{Bax016} G.~J. Baxter, D. Cellai, S.~N. Dorogovtsev, A.~V. Goltsev, J.~F.~F. Mendes. A Unified Approach to Percolation Processes on Multiplex Networks. In \textit{Interconnected Networks}, Ed.\ A. Garas, 101--123 (Springer International Publishing, Berlin, Germany, 2016).
\bibitem{Men012} A. Saumell-Mendiola, M.~Á. Serrano, M. Boguñá. Epidemic spreading on interconnected networks. \textit{Phys.\ Rev.\ E} \textbf{86}, 026106 (2012).
\bibitem{Gran013} C. Granell, S. G\'omez, A. Arenas. Dynamical Interplay between Awareness and Epidemic Spreading in Multiplex Networks. \textit{Phys. Rev. Lett.} \textbf{111}, 128701 (2013).
\bibitem{Buono014} C. Buono , L.G. Alvarez-Zuzek, P.A. Macri, L.A. Braunstein. Epidemics in Partially Overlapped Multiplex Networks. \textit{PLoS One} \textbf{9}, e92200 (2014).
\bibitem{Sanz014} J. Sanz, Ch.-Y. Xia, S. Meloni, Y. Moreno. Dynamics of Interacting Diseases. \textit{Phys. Rev. X} \textbf{4}, 041005 (2014).
\bibitem{MAB016} G. Menichetti, L. Dall'Asta, G. Bianconi. Control of Multilayer Networks. \textit{Sci.\ Rep.} \textbf{6}, 20706 (2016).
\bibitem{JGG012} J. G\'omez-Garde\~nes, I. Reinares, A. Arenas, L.M. Flor\'{\i}a. Evolution of Cooperation in Multiplex Networks. \textit{Sci. Rep.} \textbf{2}, 620 (2012).
\bibitem{Wang014} Z. Wang, A. Szolnoki, M. Perc. Rewarding evolutionary fitness with links between populations promotes cooperation. \textit{J. Theor. Biol.} \textbf{349}, 50 (2014).
\bibitem{Mata015} J.~T. Matamalas, J. Poncela-Casasnovas, S. G\'omez, A. Arenas. Strategical incoherence regulates cooperation in social dilemmas on multiplex networks \textit{Sci. Rep.} \textbf{5}, 9519 (2015).
\bibitem{Wang015} Z. Wang, L. Wang, A. Szolnoki, M. Perc. Evolutionary games on multilayer networks: a colloquium. \textit{Eur. Phys. J. B} \textbf{88}, 124 (2015)
\bibitem{Gom013} S. Gómez, A. Díaz-Guilera, J. Gómez-Gardeñes, C.~J. Pérez-Vicente, Y. Moreno, A. Arenas. Diffusion dynamics on multiplex networks. \textit{Phys.\ Rev.\ Lett.} \textbf{110}, 028701 (2013).
\bibitem{Alb000} R. Albert, H. Jeong, A.-L. Barabási, Error and attack tolerance of complex networks, \textit{Nature} \textbf{406}, 378--382 (2000).
\bibitem{Dan016} M.~M. Danziger, L.~M. Shekhtman, A. Bashan, Y. Berezin, S. Havlin. Vulnerability of Interdependent Networks and Networks of Networks. In \textit{Interconnected Networks}, Ed.\ A. Garas, 79--99 (Springer International Publishing, Berlin, Germany, 2016).
\bibitem{Agu015} J. Aguirre, R. Sevilla-Escoboza, R. Gutiérrez, D. Papo, and J.~M. Buldú. Synchronization of interconnected networks: the role of connector nodes. \textit{Phys.\ Rev.\ Lett.} \textbf{112}, 248701 (2015).
\bibitem{Zha015} X. Zhang, S. Boccaletti, S. Guan, Z. Liu. Explosive synchronization in adaptive and multilayer networks. \textit{Phys.\ Rev.\ Lett.} \textbf{114}, 038701 (2015).
\bibitem{Sev015} R. Sevilla-Escoboza, R. Gutiérrez, G. Huerta-Cuellar, S. Boccaletti, J. Gómez-Gardeñes, A. Arenas, and J.~M. Buldú. Enhancing the stability of the synchronization of multivariable coupled oscillators. \textit{Phys.\ Rev.~E} \textbf{92}, 032804 (2015).
\bibitem{Gambuzza15} L.~V. Gambuzza, M. Frasca, J. G\'omez-Garde\~nes. Intra-layer synchronization in multiplex networks. \textit{EPL} \textbf{110}, 20010 (2015).
\bibitem{Sor012} F. Sorrentino. Synchronization of hypernetworks of coupled dynamical systems. \textit{New J. Phys.} \textbf{14}, 033035 (2012).
\bibitem{Irv012} D. Irving, F. Sorrentino. Synchronization of dynamical hypernetworks: Dimensionality reduction through simultaneous block-diagonalization of matrices. \textit{Phys.\ Rev.~E} \textbf{86}, 056102 (2012).
\bibitem{Gom006} J. Gómez-Gardeñes, Y. Moreno. From scale-free to Erd\H{o}s-Rényi networks. \textit{Phys.\ Rev.~E} \textbf{73}, 056124 (2006).
\bibitem{design1} L. Donetti, P. Hurtado, M.A. Mu\~noz. Entangled networks, synchronization, and optimal network topology. {\em Phys. Rev. Lett.} \textbf{95}, 188701 (2005).
\bibitem{design2} T. Nishikawa, A.E. Motter. Synchronization is optimal in non-diagonalizable networks. \textit{Phys. Rev. E} \textbf{73}, 065106 (2006).
\bibitem{design3} C. Zhou, J. Kurths. Dynamical Weights and Enhanced Synchronization in Adaptive Complex Networks. \textit{Phys. Rev. Lett.} \textbf{96}, 164102 (2006).
\bibitem{design4}  T. Nishikawa, A.E. Motter. Network synchronization landscape reveals compensatory structures, quantization, and the positive effect of negative interactions. \textit{Proc. Nat. Acad. Sci USA} \textbf{107}, 10342--10347 (2010).
\bibitem{Nepusz012} T. Nepusz, T. Vicsek, Controlling edge dynamics in complex networks. \textit{Nat. Phys.} \textbf{8}, 568--573 (2012).
\bibitem{Sun013} J. Sun, A. E. Motter, Controllability transition and nonlocality in network control. \textit{Phys. Rev. Lett.} \textbf{110}, 208701 (2013).
\bibitem{Gao014} J. Gao, Y.-Y. Liu, R. M. D'Souza, A.-L. Barab\'asi, Target control of complex networks. \textit{Nat. Commun.} \textbf{5}, 5415 (2014).
\bibitem{Skardal015} P.~S. Skardal, A. Arenas, Control of coupled oscillator networks with application to microgrid technologies. \textit{Science Advances} \textbf{1}, e1500339 (2015).
\bibitem{Ska14} P.~S. Skardal, D. Taylor, J. Sun. Optimal Synchronization of Complex Networks. \textit{Phys.\ Rev.\ Lett.} \textbf{113}, 144101 (2014).
\bibitem{Ska16} P.~S. Skardal, D. Taylor, J. Sun. Optimal synchronization of directed complex networks. \textit{Chaos} \textbf{26}, 094807 (2016).
\end{thebibliography}
\end{document}